\documentclass[prr, twocolumn, groupedaddress, showpacs, preprintnumbers, amsmath, amssymb, aps, floatfix, superscriptaddress]{revtex4-2}

\usepackage{siunitx}
\usepackage{graphicx}
\usepackage{verbatim}
\usepackage{dcolumn}
\usepackage[ruled,vlined]{algorithm2e}
\usepackage{color}
\usepackage{xcolor}
\usepackage[colorlinks = true, linkcolor = blue, urlcolor  = blue, citecolor = blue, anchorcolor = blue]{hyperref}
\usepackage{mathtools}
\usepackage{orcidlink}

\DeclareMathOperator*{\argmin}{arg\,min}

\begin{document}


\title{Coupled Twiss Parameters Estimation from Turn-by-Turn Data}


\author{I.~Morozov\,\orcidlink{0000-0002-1821-7051}}
\email{i.a.morozov@inp.nsk.su}
\affiliation{Budker Institute of Nuclear Physics SB RAS, Novosibirsk 630090, Russia}
\affiliation{SRF Siberian Circular Photon Source "SKIF" Boreskov Institute of Catalysis SB RAS, Koltsovo 630559, Russia}
\affiliation{Novosibirsk State Technical University, Novosibirsk 630073, Russia}
\author{Yu.~Maltseva\,\orcidlink{0000-0003-2753-2552}}
\affiliation{Budker Institute of Nuclear Physics SB RAS, Novosibirsk 630090, Russia}
\affiliation{Novosibirsk State Technical University, Novosibirsk 630073, Russia}
\date{\today}




\begin{abstract}

Linear optics parameters are often estimated using turn-by-turn (TbT) data.
In-plane beta functions, which are the most common measurement objectives, can be estimated from the amplitudes or phases of TbT data, providing an overall characterization of the linear lattice.
In addition to estimating uncoupled Twiss parameters, TbT data can also be utilized for characterizing coupled linear motion.
We investigate several methods for constructing a full normalization matrix at each beam position monitor (BPM). 
BPMs provide information on beam centroid transverse coordinates, but direct observation of transverse momenta is not available.
To estimate momenta, one can use a pair of BPMs or fit data obtained from several BPMs. 
By using both coordinates and momenta, it is possible to fit the one-turn matrix (or its power) at each BPM, allowing for the computation of coupled Twiss parameters.
Another approach  involves the minimization of side peak amplitudes in the spectrum of normalized complex coordinates.
Similarly, the normalization matrix can be estimated by fitting linear coupled invariants.
In this study, we derive and utilize a special form of the normalization matrix, which remains non-singular in the zero coupling limit.
Coupled Twiss parameters can be obtained from the normalization matrix.
The paper presents the results of applying and comparing these methods to both modeled and measured VEPP-4M TbT data, demonstrating effective estimation of coupled Twiss parameters.

\end{abstract}


\maketitle


\section{Introduction}
\label{sec:introduction}


The accurate measurement of accelerator linear optics parameters is crucial for achieving optimal performance in particle accelerators.
In Tom\'as et al.~\cite{tomas} a comprehensive overview of different methods for measuring and correcting the optical model of an accelerator is presented.
One of the most common methods for linear optics measurement and correction is based on the orbit response matrix (ORM)~\cite{response} and usually used in light sources such as SIRIUS, MAX-IV, and ALS-U~\cite{sirius, max, alsu}.
Despite the high accuracy of this method in the estimation of the linear model and its correction, it has certain drawbacks, such as it takes a long time to measure the ORM and the potential for overfitting problem~\cite{tomas}.
It should be noted that several variations of this method exist~\cite{inv_response, ac_response, better_ac_response}, primarily focused on reducing the measurement time for the ORM. 


Methods utilizing TbT data are also widely used for the determination of uncoupled optical functions~\cite{castro, nbpm, better_nbpm}.
Although not always as accurate as the methods based on the ORM, like in the ESRF case~\cite{franchi}, these methods provide fast optics inference. In these methods, optical functions can be estimated using the amplitude or phase of coherent oscillations derived from the harmonic analysis of TbT signals.
Typically, the ideal linear model is designed to be uncoupled, and the corresponding optical functions are estimated without coupling. 
However, estimating coupled linear Twiss parameters is also important, especially when the initial accelerator model includes coupling by design or when the lattice contains sources of undesired coupling that need identification.
For example, in~\cite{hofer} effects of undesired coupling are discussed for the LHC. The SKIF light source~\cite{skif}, currently under construction, will feature a special operation mode that uses coupling to increase the beam lifetime~\cite{baranov}.
Similar to methods based on the closed orbit response, methods based on TbT signals are applicable for examining the coupled optical model~\cite{wolski, sagan2000, robust, calaga2005}.


In this paper, we consider several methods for estimating the linear coupled Twiss parameters of transverse motion.
To achieve this, we construct the normalization matrix at each BPM location, transforming the original phase space coordinates into normalized ones.
In this normalized space, the motion is uncoupled, and the phase space trajectories are circular.
BPMs provide information only about the beam centroid coordinates, but not the momenta.
We explore several methods for reconstructing transverse momenta in a linear approximation~\cite{berz, ipac2016, wegscheider}.
With both coordinates and momenta, it is feasible to approximate the one-turn matrix at each BPM~\cite{robust}.
For each fitted matrix, we perform a symplectification procedure~\cite{symplectic} and determine the corresponding parameters of the coupled optical model~\cite{wolski}.
It is noteworthy that Twiss parameters can be computed by fitting not just the one-turn matrix but also its power, potentially enhancing the signal-to-noise ratio~\cite{robust}.
Further examples of linear coupled motion parameterizations are discussed in references~\cite{et, rip, leb, sagan1999}.


In addition to the method based on the one-turn matrix, this paper introduces two more methods for estimating coupled Twiss parameters. 
These methods involve approximating normalization matrices using spectra of complex normalized coordinates and linear coupled invariants.
They use a special form of the normalization matrix that does not contain singularities in the limit of zero coupling.
All the methods discussed in this article rely on the computation of reconstructed transverse momenta using two or more BPMs. 
This makes them dependent on the BPM calibration accuracy.


The investigated methods for estimating coupled Twiss parameters have been successfully tested through numerical simulations.
These tests demonstrated high accuracy, even in the presence of random errors in TbT signals and lattice focusing errors.
When applied to experimental data from the VEPP-4M accelerator~\cite{vepp4m}, these methods also yielded promising results.
They showed consistency with established uncoupled methods that rely on harmonic analysis of TbT signals~\cite{rupac2021_anomaly, rupac2021_twiss}.
The resulting normalization matrices convert phase trajectories in original coordinates into circles in normalized coordinates.


Other methods, such as fitting coefficients of the coupled invariant from TbT data~\cite{li} and obtaining coupled Twiss parameters using only the phases of TbT signals~\cite{wolski}, provide alternative approaches for determining the coupled linear Twiss parameters. 


The structure of this article is as follows:
Section~\ref{sec:upcoupled} briefly reviews the measurement of uncoupled Twiss parameters from TbT signals.
Section~\ref{sec:coupled} describes the parameterization of linear coupled motion, as introduced in~\cite{wolski}.
The procedure for explicit construction of the normalization matrix based on the symplectic condition is presented in Section~\ref{sec:normal}.
Reconstruction of transverse momenta is described in Section~\ref{sec:momenta}.
Sections \ref{sec:matrix}, \ref{sec:spectrum}, and \ref{sec:invariant} introduce methods for estimating coupled Twiss parameters based on the one-turn matrix, the complex normalized spectrum, and coupled linear invariants, respectively.
The experimental results for the VEPP-4M accelerator are presented in Section~\ref{sec:vepp}.


\section{Uncoupled Twiss Parameters}
\label{sec:upcoupled}


The modern BPM systems enable the measurement of transverse TbT coordinates with high accuracy. 
If coherent oscillations around the closed orbit are excited, for example, by a pulsed electromagnetic field, then the signal from BPM $i$ can be used to estimate the frequency $\nu$, amplitude $a_i$, and phase $\varphi_i$ of the oscillations. 
In the linear uncoupled case, oscillations around the closed orbit can be described as follows:
\begin{align}
  q_i(n) &= a_i \cos(2 \pi \nu n + \varphi_i) \nonumber \\ 
         &=  \sqrt{2 I \beta_i} \cos(2 \pi \nu n + \varphi_i)    
\end{align}
where $I$ is the linear action, and $\beta_i$ is the beta function at BPM $i$.
Measured TbT amplitude values can be used to estimate $\beta_i$ value, provided the value of the action $I$ is known.
The action itself can be estimated using either model values or values measured by a different method of the beta function $\bar \beta_i$, and measured amplitude values as an average value over all BPMs:
\begin{equation}
    \hat I = \frac{1}{2}{\left<\frac{a_i^2}{\bar \beta_i}\right>}_i
\end{equation}
The beta functions at each BPM can be estimated using approximated action $\hat I$ as follows:
\begin{equation}
    \beta_i = \frac{a_i^2}{2 \hat I}
\end{equation}


This method allows for a fast measument of beta-functions and does not require synchronization of BPMs. However, it relies on the calibration of BPMs and the accuracy of linear action estimation.
Significant deviations in the lattice from the model can lead to inaccuracies in the estimated action, potentially causing biases in the beta functions.
The action estimate can be improved by using not model, but measured values of the beta function.
For instance, it is possible to use the estimates derived using phases of TbT signals.


Amplitudes allow the estimation of only beta functions, which are of primary interest but do not provide a complete description of the motion.
 The Courant-Snyder parameters $\alpha$ and $\beta$ can be used for the full parameterization of linear uncoupled motion.
The transition matrix between locations $i$ and $j$ can be represented using these parameters as:
\begin{equation}
\label{normal_transform}
    M_{i, j} = N_{j} R_{i, j}(\mu_{i, j}) N_{i}^{-1}
\end{equation}
where $R_{i, j}(\mu_{i, j})$ is the rotation matrix with angle $\mu_{i, j}$, representing the phase advance between locations $i$ and $j$, and $N_{i}$ and $N_{j}$ are normalization matrices corresponding to locations $i$ and $j$, respectively.


\begin{figure}[!h]
    \centering
    \includegraphics[width=1.0\columnwidth,height=0.25\columnwidth]{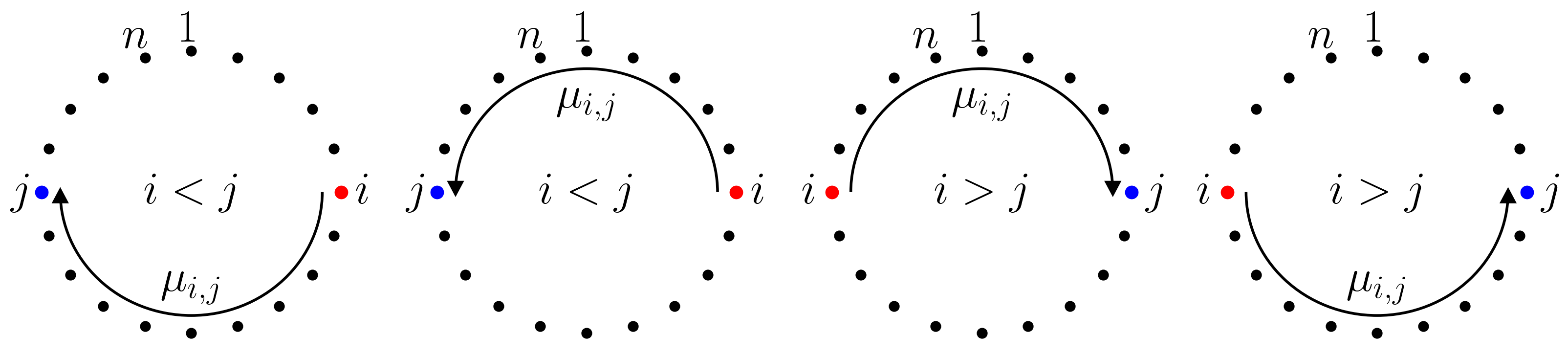}
    \caption{ (Color) Different cases of phase advance computation from $i$ to $j$. Phase advance is considered negative if propagation from $i$ to $j$ implies moving back in time. If BPMs are located on different turns, the tune should be adjusted before computing the phase advance.}
    \label{fig:circle}
\end{figure}


Figure~\ref{fig:circle} illustrates the computation of phase advance from location $i$ to $j$, given the total number of BPMs $n$, where $i$ and $j$ can be any integers.
Given a location index $i \in \mathbb{Z}$, its corresponding monitor index is defined as $\bar i = i \mod n$, i.e. index on the first turn.
With the measured phase $\varphi_{\bar i}$ at this index and a frequency $\nu$, the phase is given by:
\begin{equation}
    \varphi_{i}=\mod\left( \varphi_{\bar i} + 2 \pi \nu k, 2 \pi \right), \quad k = \frac{i - \bar i}{n}
\end{equation}
Using the above expression, the phase advance from location $i$ to $j$ is expressed as:
\begin{equation}
    \mu_{i, j} = (-1)^{i > j} \mod \left((-1)^{i < j} \left( \varphi_{i} - \varphi_{j} \right), 2 \pi\right)
\end{equation}
where the expressions $i > j$ and $i < j$ are logical conditions, and the function $\mod(x, y)$ returns the remainder of the division of $x$ by $y$.


The normalization matrix $N$ is symplectic and depends on the Twiss parameters as follows:
\begin{equation}
\label{norma_cs}
    N = \begin{bmatrix}\sqrt{\beta} & 0 \\ -\alpha/ \sqrt{\beta} & 1 / \sqrt{\beta}\end{bmatrix}
\end{equation}
where the element $n_{1, 1} = \sqrt{\beta}$ is positive, and the element $n_{1, 2}$ is identically equal to zero. 
This is a standard gauge used to set the form of $N$ so that the beam size depends only on the beta function.

For this particular form of the normalization matrix (Equation~\ref{norma_cs}), given monitors $i$ and $j$, and the transport matrix $M_{i, j}$ between them, the following identity holds:
\begin{widetext}
\begin{equation}
\label{equation:normal}
0 \equiv n^{j}_{1, 2} = \left(
\begin{bmatrix} m^{(i, j)}_{1, 1} & m^{(i, j)}_{1, 2} \\ m^{(i, j)}_{2, 1} & m^{(i, j)}_{2, 2} \end{bmatrix}
\begin{bmatrix}\sqrt{\beta_i} & 0 \\ -\alpha_i / \sqrt{\beta_i} & 1 / \sqrt{\beta_i} \end{bmatrix}
\begin{bmatrix} \cos{\mu_{i, j}} & -\sin{\mu_{i, j}} \\ \sin{\mu_{i, j}} & \cos{\mu_{i, j}} \end{bmatrix}
\right)_{1, 2}
\end{equation}
\end{widetext}
where $m^{(i, j)}_{q, p}$ are the elements of the transport matrix $M_{i, j}$, $\alpha_i$ and $\beta_i$ are the uncoupled Twiss parameters at location $i$, and $\mu_{i, j}$ is the phase advance form $i$ to $j$.


The identity above provides an equation that relates Twiss parameters at location $i$. 
To determine parameters $\alpha_i$ and $\beta_i$ from Equation~\ref{equation:normal}, at least two additional monitors, apart from the probed one, are required. 
This requirement forms the basis of the well-known triplet method, as defined in~\cite{castro}.
Model transport matrices can be utilized to estimate measured Twiss parameters.
It should be noted that the measurable values of $\alpha$ and $\beta$ are conveniently expressed through the measured phase advance, model values $\alpha^{m}$ and $\beta^{m}$, and model phase advance as follows:
\begin{widetext}
    \begin{align}
    \label{equation:beta}
        &\alpha_i=\alpha^{m}_i\frac{\cot(\mu_{ij})-\cot(\mu_{ik})}{\cot(\mu^{m}_{ij})-\cot(\mu^{m}_{ik})} -\cot(\mu_{ij}) \sec(\mu^{m}_{ij} - \mu^{m}_{ik})\cos(\mu^{m}_{ik})\sin(\mu^{m}_{ij}) + \cot(\mu_{ik})\sec(\mu^{m}_{ij}-\mu^{m}_{ik})\cos(\mu^{m}_{ij})\sin(\mu^{m}_{ik}) \nonumber \\ 
        &\beta_i=\beta^{m}_i\frac{\cot(\mu_{ij})-\cot(\mu_{ik})}{\cot(\mu^{m}_{ij})-\cot(\mu^{m}_{ik})}
    \end{align}
\end{widetext}
where $\alpha^{m}_i$ and $\beta^{m}_i$ are the model Twiss parameters at the probed monitor $i$, $\mu^{m}_{ij}$ and $\mu^{m}_{ik}$ are the model phase advances from $i$ to $j$ and $k$, respectively, while $\mu_{ij}$ and $\mu_{ik}$ are the corresponding measured phase advances.


Since the method for estimating Twiss parameters relies solely on phase measurements, it is independent of BPM calibrations.
This even allows investigating the calibrations of the monitors themselves~\cite{cal}. 
The method can be further improved ~\cite{nbpm, better_nbpm}.
To improve estimation accuracy, the approach employs several triples of monitors, taking into consideration the impact of systematic errors in the sections between these monitors. 
If both transverse planes are considered simultaneously, one matrix equation yields two scalar equations for the elements of the normalization matrix.
Furthermore, when analyzing both transverse planes simultaneously, as is the case in coupled systems, the matrix equation yields two scalar equations for normalization matrix elements.


At the VEPP-4M accelerator, the estimation of Twiss parameters for the linear optical model is performed using methods based on both amplitude~\cite{smaluk} and phase~\cite{rupac2021_twiss}.
TbT signals are collected from 54 BPMs, which, in this mode of operation, have a resolution close to \SI{10}{\micro\metre}\cite{beh}. 
Measured TbT signals are also checked for anomalous spikes and synchronization errors~\cite{rupac2021_anomaly}.
For the phase based measurement method, selection and averaging with weights accounting for random measurement errors are performed for several triples of monitors.


\section{Coupled Twiss Parameters}
\label{sec:coupled}


Different parameterizations, as referenced in~\cite{et, rip, leb, sagan1999}, are used to transform linear coupled motion into its normal form.
While most parameterizations primarily focus on the transverse planes, this paper utilizes the parameterization proposed by Wolski~\cite{wolski}.
The main objective of any parameterization is to construct a symplectic transformation. This transformation converts the original system, which includes coupling, into a normalized uncoupled system, where the phase trajectories form circles:
\begin{equation}
    X_{n + 1} = R X_{n} = N^{-1} M N X_{n}
\end{equation}
where $R$ is the rotation matrix corresponding to the normalized system, $M$ is the one-turn matrix of the original system, $N$ is the normalization matrix.
$X_{m}$ and $X_{m+1}$ are the normalized coordinates at turn $n$ and $n+1$, respectively.
The matrix $N$ must be symplectic and comply with the chosen gauge. For instance, the condition $n_{i, i + 1} \equiv 0,~i = 2 k + 1$ and $n_{i, i} > 0$ can be applied to establish a direct relationship between the $n_{i, i}$ elements and the beam sizes.


In Wolski's paper, the normalization matrix $N$ is derived from the eigenvectors of the one-turn matrix $M$:
\begin{equation}
    M e_{\pm k} = \lambda_{\pm k} e_{\pm k} 
\end{equation}
where the eigenvectors $e_{k}$ are normalized such that $e_{k}^{H} S e_{k'} = i \delta_{k, k'}$ where $H$ denotes 
the Hermitian transpose and symplectic matrix $S$ is given by:
\begin{equation}
\label{equation:sm}
S = \begin{pmatrix*}[r]
0 & 1 & 0 & 0 \\
-1 & 0 & 0 & 0 \\
0 & 0 & 0 & 1  \\
0 & 0 & -1 & 0 
\end{pmatrix*}
\end{equation}
It is important to note that the one-turn matrix $M$ and the matrix $\Sigma S$ (where $\Sigma$ is a transformation matrix for beam second moments) share the same eigenvectors. The transformation of beam second moments is governed by the equation:
\begin{equation}
    \Sigma_j = M_{i, j} \Sigma_i M_{i, j}^{T}
\end{equation}
where $M_{i, j}$ is a transport matrix between $i$ and $j$.


The normalizing matrix $N$ is constructed from the eigenvectors of the one-turn matrix $M$ as follows:
\begin{equation}
    N = E Q = \begin{pmatrix} e_{-I} & e_{I} & e_{-II} & e_{II} \end{pmatrix} \\
    \begin{pmatrix} \hat Q & 0 \\ 0 & \hat Q \end{pmatrix}
\end{equation}
and the $\hat Q$ matrix is constructed from eigenvectors of rotation matrix and is given by:
\begin{equation}    
    \hat Q = \frac{1}{\sqrt{2}}\begin{pmatrix} 1 & i \\ 1 & -i\end{pmatrix}
\end{equation}
Here, the eigenvectors $e_{-I}$ and $e_{I}$ correspond to the first degree of freedom, while $e_{-II}$ and $e_{II}$ correspond to the second degree of freedom.
For each degree of freedom $k$, Twiss matrices $B^{k}$ are introduced:
\begin{equation}
    B^{k} = N T^{k} N^{T}
\end{equation}
where matrices $T^{k}$ perform the projection, for example, for two degrees of freedom:
\begin{equation}
T^{I} = \begin{pmatrix}
1 & 0 & 0 & 0 \\
0 & 1 & 0 & 0 \\
0 & 0 & 0 & 0  \\
0 & 0 & 0 & 0 
\end{pmatrix}, \quad
T^{II} = \begin{pmatrix}
0 & 0 & 0 & 0 \\
0 & 0 & 0 & 0 \\
0 & 0 & 1 & 0  \\
0 & 0 & 0 & 1 
\end{pmatrix}
\end{equation}


The Twiss parameter matrices $B^{k}$ are symmetric by construction and are transformed similarly to the matrix of the second moments $\Sigma$:
\begin{equation}
\label{twiss_transform}
    B^{k}_j = M_{i, j} B^{k}_i M_{i, j}^{T}
\end{equation}


This parameterization method, while being redundant, provides a convenient form of Twiss parameters that can be easily transformed from one location to another. 
Additionally, it is applicable in any dimension. In this paper, we use a software implementation of this normalization procedure~\cite{package}. 
It is also possible to calculate higher order derivatives of the Twiss parameters with respect to different parameters~\cite{torch, ndmap}.


In Wolski's article, a method for estimating coupled Twiss parameters from phase data is presented.
Equations similar to~\ref{equation:normal} in Section~\ref{sec:upcoupled} are also valid in the coupled case with a modified normalization matrix $N_i$ and with full $4 \times 4$ transport matrix $M_{i,j}$ between monitors. 
Thus, is it possible to estimate coupled Twiss parameters using a larger number of BPMs.
However, this method has several limitations. 
One of them is the requirement to use a large area around the probed monitor.
Additionally, model transport matrices typically do not contain coupling  by design, which results in the inability to compute coupling-specific Twiss parameters.


\section{Parametric Normalization Matrix}
\label{sec:normal}


Apart from the coupled motion parameterization suggested by Wolski, we also use a normalization matrix that is constructed directly using the standard gauge and the symplectic condition.
This matrix is used to parameterize the transverse coupled motion, and its free elements are referred to as the coupled Twiss parameters.
The free elements of this matrix are selected such that the expressions for all other elements, which are fixed by the symplectic condition, do not contain singularities in the limit of zero coupling.
This is useful, since the accelerator model normally does not include coupling by design.
Several such direct parameterizations are possible for $4 \times 4$ normalization matrices.


The symplecticity condition reduces the number of free elements of $4 \times 4$ matrix to $10$, and $2$ elements, $n_{1, 2}$ and $n_{3,4}$, are fixed to zero by the standard gauge.
Additionally, the conditions $n_{1, 1} > 0$ and $n_{3, 3} > 0$ are assumed.
The normalization matrix then has the following form:
\begin{equation}
N = \begin{pmatrix*}[r]
n_{1,1} & 0 & n_{1,3} & n_{1,4} \\
n_{2,1} & X & X & X \\
n_{3,1} & X & n_{3,3} & 0  \\
n_{4,1} & X & n_{4,3} & X
\end{pmatrix*}
\end{equation}
where the parameters fixed by symplectic condition are denoted by the symbol $X$ and the elements $n_{1, 1} > 0$, $n_{1, 2} \equiv 0$, $n_{1,3}$, $n_{1,4}$, $n_{2,1}$, $n_{3,1}$, $n_{3,3} > 0$, $n_{3,4} \equiv 0$, $n_{4,1}$, and $n_{4,3}$, other elements are determined by solving algebraic equations obtained from symplectic condition $N^{T} S N = S$ with $S$ given by Equation~\ref{equation:sm} in Section~\ref{sec:coupled}.
This parametric normalization matrix can be used to relate linear invariants in measured and normalized coordinates.
Different forms of coupled Twiss parameters can be computed from normalization matrix.
The free elements for a particular normalization matrix used in this paper are given by:
\begin{widetext}
\begin{align}
& n_{2,2} = n_{3,3} (n_{1,1} + n_{1,4} (n_{3,3} n_{4,1} - n_{3,1} n_{4,3}))/(n_{1,1} (n_{1,1} n_{3,3} - n_{1,3} n_{3,1})) \nonumber \\
& n_{2,3} = (n_{1,3} n_{2,1} + n_{3,3} n_{4,1} - n_{3,1} n_{4,3})/n_{1,1} \nonumber \\
& n_{2,4} = (n_{1,4} n_{2,1} n_{3,3} - n_{3,1} + n_{1,4} n_{3,1}/n_{1,1} (n_{3,1} n_{4,3} - n_{1,3} n_{2,1} - n_{3,3} n_{4,1}))/(n_{1,1} n_{3,3} - n_{1,3} n_{3,1}) \nonumber \\
& n_{3,2} = n_{1,4} n_{3,3}/n_{1,1} \nonumber \\
& n_{4,2} = (n_{1,3} (-1 - (n_{1,4} n_{3,3} n_{4,1})/n_{1,1}) + n_{1,4} n_{3,3} n_{4,3})/(n_{1,1} n_{3,3} - n_{1,3} n_{3,1}) \nonumber \\
& n_{4,4} = (n_{1,1} + n_{1,4} (n_{3,3} n_{4,1} - n_{3,1} n_{4,3}))/(n_{1,1} n_{3,3} - n_{1,3} n_{3,1})
\end{align}
\end{widetext}


\section{Transverse Momenta Reconstruction}
\label{sec:momenta}


BPMs do not provide direct measurements of transverse momenta.
However, values of transverse momenta are necessary for estimating the coupled Twiss parameters with the methods described in this paper. 
Given the coordinates at two monitors and the model transformation between them, one can compute the transverse momenta at these monitors.
The reconstruction procedure based on Taylor series representation of the transformation can be performed in the general nonlinear case~\cite{berz, ipac2016, package}.
In the linear case, transverse momenta can be expressed by explicit formulas. 
For instance, using the coordinates from monitors $i$ and $j$, the momenta at monitor $i$ are equal to:
\begin{widetext}
\begin{align}
\label{px}
p_{x, i}  &= q_{x,i}(m^{(i,j)}_{1,1}  m^{(i,j)}_{3,4} - m^{(i,j)}_{1,4}   m^{(i,j)}_{3,1})/(m^{(i,j)}_{1,4} m^{(i,j)}_{3,2} - m^{(i,j)}_{1,2} m^{(i,j)}_{3,4})
    + q_{x,j} m^{(i,j)}_{3,4}/(m^{(i,j)}_{1,2}   m^{(i,j)}_{3,4} - m^{(i,j)}_{1,4}   m^{(i,j)}_{3,2})  \nonumber \\
    &+ q_{y,i}(m^{(i,j)}_{1,3}  m^{(i,j)}_{3,4} - m^{(i,j)}_{1,4}   m^{(i,j)}_{3,3})/(m^{(i,j)}_{1,4} m^{(i,j)}_{3,2} - m^{(i,j)}_{1,2} m^{(i,j)}_{3,4})
    + q_{y,j} m^{(i,j)}_{1,4}/(m^{(i,j)}_{1,4}   m^{(i,j)}_{3,2} - m^{(i,j)}_{1,2}   m^{(i,j)}_{3,4}) 
\end{align}
\begin{align}
\label{py}
p_{y,i}  &= q_{x,i} (m^{(i,j)}_{1,1}  m^{(i,j)}_{3,2} - m^{(i,j)}_{1,2}   m^{(i,j)}_{3,1})/(m^{(i,j)}_{1,2} m^{(i,j)}_{3,4} - m^{(i,j)}_{1,4} m^{(i,j)}_{3,2})
    + q_{x,j}  m^{(i,j)}_{3,2}/(m^{(i,j)}_{1,4}   m^{(i,j)}_{3,2} - m^{(i,j)}_{1,2}   m^{(i,j)}_{3,4})\nonumber \\
    &+ q_{y,i} (m^{(i,j)}_{1,2}  m^{(i,j)}_{3,3} - m^{(i,j)}_{1,3}   m^{(i,j)}_{3,2})/(m^{(i,j)}_{1,4} m^{(i,j)}_{3,2} - m^{(i,j)}_{1,2} m^{(i,j)}_{3,4})
    + q_{y,j}  m^{(i,j)}_{1,2}/(m^{(i,j)}_{1,2}   m^{(i,j)}_{3,4} - m^{(i,j)}_{1,4}   m^{(i,j)}_{3,2}) 
\end{align}
\end{widetext}
where $m^{(i, j)}_{q, p}$ are elements of the transport matrix from monitor $i$ to  $j$. 
In the above, the mean values are assumed to be subtracted from the measured coordinates, prior to momenta computation, since Equations~\ref{px} and \ref{py} are valid around the closed orbit.


Given a pair of BPMs along with a model or a measured transport matrix between them, it is possible to reconstruct the values of transverse momenta. 
To improve the accuracy of transverse momenta estimation, multiple BPM pairs can be utilized. 
Combining results from different monitor pairs helps obtaining a more accurate and reliable estimate of momenta.
Results from different pairs can be combined with weights, for example, using TbT noise estimate from signals without excitation or directly from oscillations~\cite{optimal}.
Alternatively, the least squares (LS) method can also be used for transverse momenta estimation.
With LS, both the coordinates and momenta are determined using the following system:
\begin{align*}
    & q_{x, j} = m^{(i, j)}_{1, 1} q_{x, i} + m^{(i, j)}_{1, 2} p_{x, i} + m^{(i, j)}_{1, 3} q_{y, i} + m^{(i, j)}_{1, 4} p_{y, i} \\
    & q_{y, j} = m^{(i, j)}_{3, 1} q_{x, i} + m^{(i, j)}_{3, 2} p_{x, i} + m^{(i, j)}_{3, 3} q_{y, i} + m^{(i, j)}_{3, 4} p_{y, i} 
\end{align*}
where the coordinates $(q_{x}, q_{y})$ and momenta $(p_{x}, p_{y})$ indexed by $i$ are fitted using several monitors within the selected range $j \in {(i - k, i + k)},~k \in \mathbb{N}^{+}$.


It should be noted that the reconstructed momenta depend on the calibrations of BPMs.
The linear part of BPM calibrations can be described by a calibration matrix:
\begin{equation}
\begin{pmatrix} \bar q_x \\ \bar q_y \end{pmatrix} = 
G
\begin{pmatrix} q_x \\ q_y \end{pmatrix} = \begin{pmatrix}
g_{x,x} &  g_{x,y} \\
g_{y,x} &  g_{y,y}
\end{pmatrix} \begin{pmatrix} q_x \\ q_y \end{pmatrix}
\end{equation}
where $g_{x,x}$ and $g_{y,y}$ are the calibration scales, $g_{x,y}$ and $g_{y,x}$ are calibration cross-talks, $q_x$ and $q_y$ correspond to actual coordinates, and $\bar q_x$ and $\bar q_y$ correspond to measured coordinates at a BPM.
Following the approach outlined in~\cite{irwin}, calibrations can be included as additional linear transformations at the BPM's entrance and exit.
The symplectic calibration matrix that transforms both coordinates and momenta is given by:
\begin{equation}
\bar G = \begin{pmatrix}
 g_{x, x} & g_{x, y} & 0 & 0 \\
 0 & 0 & g_{y, y} / \det G & -g_{y, x} / \det G \\
 g_{y, x} & g_{y, y} & 0 & 0 \\
 0 & 0 & -g_{x, y} / \det G & g_{x, x} / \det G
\end{pmatrix}
\end{equation}
Consequently, the observed transport matrix between BPMs $i$ and $j$ is given by:
\begin{equation}
    \bar M_{i, j} = \bar G_j M_{i, j} \bar G_i^{-1}
\end{equation}
This illustrates how calibration errors can modify the observed twiss parameters and, hence, the importance of calibration independent methods based on phase measurements.
Since methods based on reconstructed momenta depend on BPM calibrations, the estimated Twiss parameters correspond to the parameters observed at the BPM frame. Transformation between BPM frame parameters and beam frame parameters can be performed using Equation~\ref{normal_transform}, or \ref{twiss_transform}.


To account for uncertainties in transverse momenta estimation and to improve statistical reliability, both the bootstrap procedure and the Monte Carlo (MC) method can be used.
The bootstrap procedure enables the assessment of random coordinate measurement errors by generating multiple samples (with replacement) from the available data.
This approach provides a distribution of possible momentum values and an estimate of its mean value and corresponding standard errors.
The MC method accounts for systematic errors in transport matrices by simulating random variations in accelerator elements and BPM parameters.
Both methods, bootstrap and MC, are powerful tools for accounting for various sources of uncertainty in estimating transverse momenta.
In the case of VEPP-4M experimental data, only the bootstrap method is applied to momenta or coupled Twiss estimation procedures.


\section{One-turn Matrix Reconstruction}
\label{sec:matrix}


Coupled Twiss parameters can be estimated from the one-turn matrix at selected BPM or its power $M^k = N R^k N^{-1}$.
Since the coordinates and momenta at different turns at a BPM are related by:
\begin{equation}
    {(q_x, p_x, q_y, p_y)}^{T}_{n + k} = M^{k} {(q_x, p_x, q_y, p_y)}^{T}_{n}
\end{equation}
the LS method can be used to solve this matrix equation for $M^{k}$.
TbT data from one or multiple measurements can be utilized to estimate  $M^{k}$.
The selection of a specific matrix power is aimed at optimizing the signal-to-noise ratio, particularly in cases of beatings associated with coupled transverse oscillations~\cite{robust}.


The estimated matrix $\hat M^{k}$ is generally not symplectic.
Therefore, symplectification~\cite{symplectic} is performed before calculating the Twiss parameters:
\begin{align}
& V = S (I - \hat M^{k}) {(\hat M^{k})}^{-1} \nonumber \\
& W = 1/2 (V + V^T) \nonumber \\
& \bar M^{k} = {(S + W)}^{-1} (S - W) 
\end{align}
where $S$ is the standard symplectic form. 
The matrix $\bar M^{k}$ serves as a symplectic approximation of the matrix $M^{k}$.
This matrix can then be used to determine the coupled Twiss parameters~\cite{wolski, package}.


To assess the uncertainties of the parameters, the bootstrap procedure is used with samples generated from TbT data.
This enables the estimation of the contribution from random measurement errors in BPM coordinates.
Some samples may yield anomalous results, significantly differing from others. Thus, when averaging across samples, a robust measure of dispersion, such as the interval between specified quartiles, is utilized to remove potential outliers.
Systematic errors arise from using either modeled or measured transport matrices for momentum calculations. However, at this stage, systematic errors are not accounted for in determining optical parameters at the VEPP-4M.


\section{Complex Normalized Spectrum}
\label{sec:spectrum}


Using the normalization matrix $N$ and coordinates $(q_x, p_x, q_y, p_y)$, one can calculate normalized coordinates as follows:
\begin{equation}
\label{frame}
    (Q_x,P_x,Q_y,P_y)^T = N^{-1} (q_x,p_x,q_y,p_y)^T
\end{equation}
In this normalized space, phase space trajectories form circles when correct normalization matrix $N$ is applied.
Introducing the following complex coordinates:
\begin{equation}
    (W_x, W_y) = (Q_x - i P_x, Q_y - i P_y)
\end{equation}
the corresponding amplitude spectra are expected to contain a single main peak at frequency $\nu_x$ for $W_x$ and at $\nu_y$ for $W_y$.
This is true, since $W_x$ and $W_y$ are eigenvectors of the rotation matrix $R(2 \pi \nu_x)$ and $R(2 \pi \nu_y)$ respectively.


\begin{figure}[ht!]
    \begin{center}
    \includegraphics[width=\columnwidth,height=0.62\columnwidth]{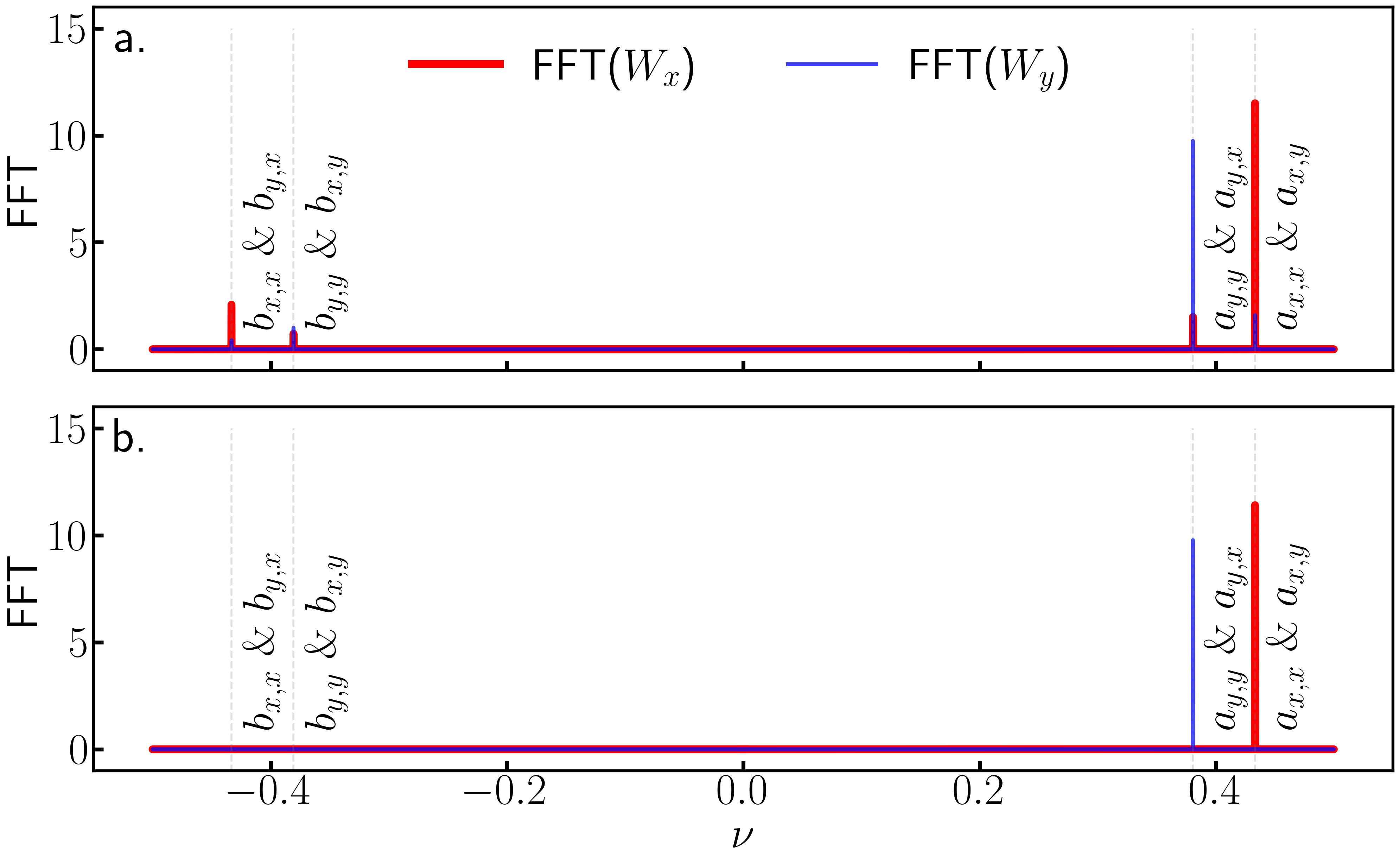}
    \end{center}
    \vspace{-0.5cm}
    \caption{
    (Color) The spectrum of oscillations for complex coordinates $W_x$ (in red) and $W_y$ (in blue) with mismatched (a.) and matched (b.) normalization matrix $N$. For the matched case, only the main amplitudes $a_{x, x}$ and $a_{y, y}$ are nonzero.}
    \label{fig:spectra}
\end{figure}


Transverse TbT signals $x$ and $y$ are available at each BPM.
The frequencies of oscillations $\nu_x$ and $\nu_y$ can be estimated with high accuracy from these signals~\cite{naff, harpy, tune}.
In the linear approximation, if $N$ does not perform full normalization, the spectrum of complex coordinates contains two peaks at frequencies $(\nu_x, \nu_y)$ and two peaks at reflected frequencies $(\bar \nu_x, \bar \nu_y) = 1 - (\nu_x, \nu_y)$. 
Figure~\ref{fig:spectra} illustrates the use of both mismatched and matched normalization matrices in computing normalized spectra.
In addition to the primary peaks at the main frequencies, additional peaks appear due to coupled oscillations for the mismatched case. 

In the case of nonlinear motion, harmonics of the main frequencies will also be present, but for measurements of linear optical model parameters, nonlinear effects can usually be neglected.
Given two main frequencies and two reflected frequencies, four amplitudes can be computed from each spectrum.
For example, the amplitudes of peaks for horizontal normalized coordinates $W_x$ can be estimated as:
\begin{align}
    & a_{x, x} = \left|\left<W_x \exp\left(i 2 \pi \nu_x n\right)\right>\right| \nonumber\\
    & b_{x, x} = \left|\left<W_x \exp\left(i 2 \pi \bar \nu_x n\right)\right>\right| \nonumber \\
    & a_{x, y} = \left|\left<W_x \exp\left(i 2 \pi \nu_y n\right)\right>\right| \nonumber \\
    & b_{x, y} = \left|\left<W_x \exp\left(i 2 \pi \bar \nu_y n\right)\right>\right|
\end{align}
where $<\dots>$ is the average over different turns $n$ and $|\dots|$ is the amplitude of a complex number.


In the above expressions the average over turns can be replaced with the Birkhoff average~\cite{tune} weighted by cosine or other window function.
A weighted average might produce more accurate parameter estimations if the noise in TbT data is relatively small.
Similarly, amplitudes $(a_{y,y}, b_{y,y}, a_{y, x}, b_{y, x})$ for vertical coordinates can be computed.


The optimization objective below can be minimized to determine the elements of the normalization matrix $N$:
\begin{equation}
    \argmin_{N}{\left(\frac{b_{x, x} + a_{x, y} + b_{x, y}}{a_{x, x}} + \frac{a_{y, x} + b_{y, x} + b_{y, y}}{a_{y, y}}\right)}
\end{equation}
The specific form of the normalization matrix, as described in Section~\ref{sec:normal}, is used in the optimization.
The optimization is performed using a nonlinear LS method with the model normalization matrix as an initial guess.
TbT coordinates, obtained from a single measurement, are used as input data.
Multiple shifted samples, having the same length but starting from different turns, are generated to assess the variability of the estimated parameters.


\section{Coupled Linear Invariants}
\label{sec:invariant}

\begin{figure}[th!]
    \begin{center}
    \includegraphics[width=\columnwidth,height=0.62\columnwidth]{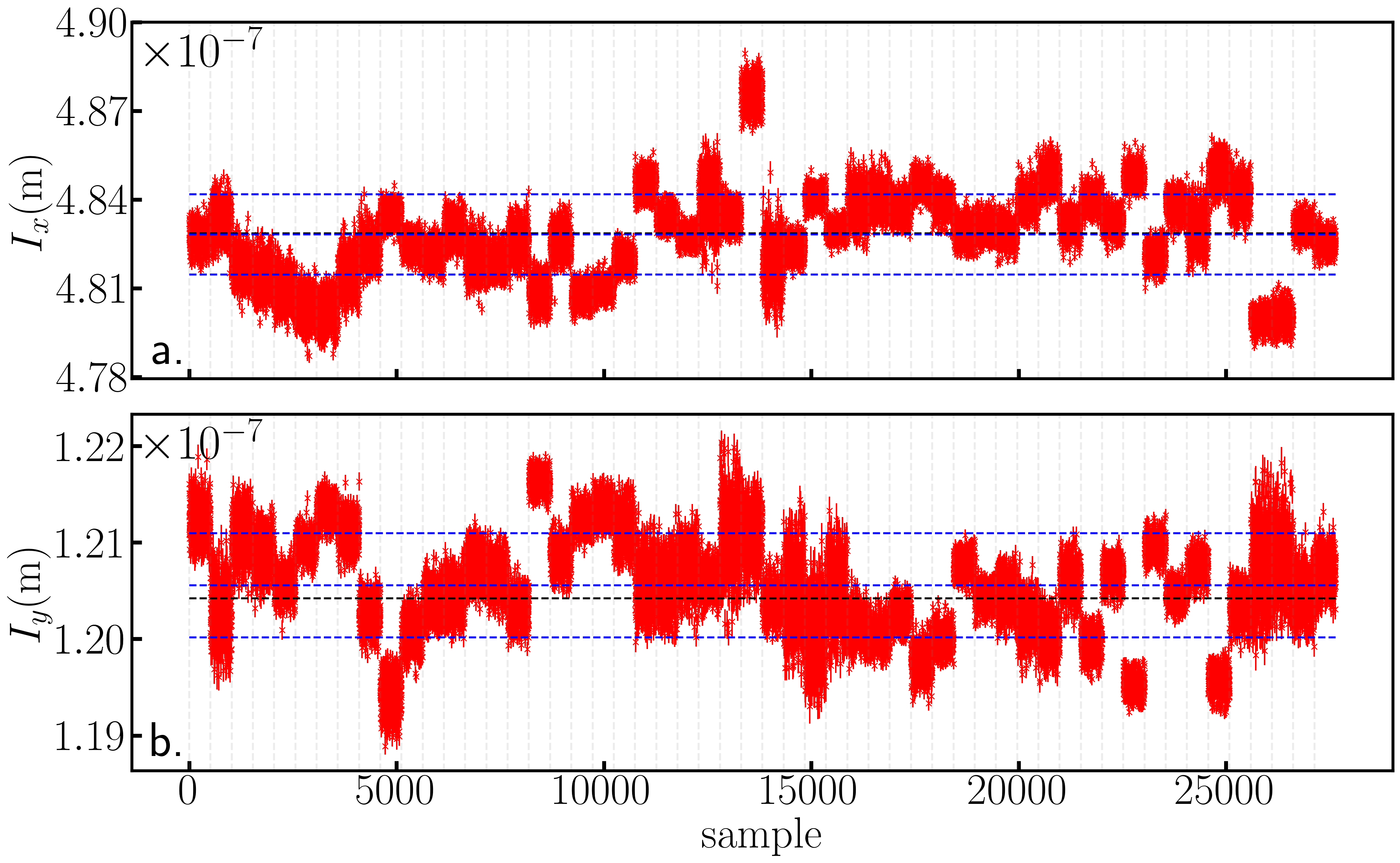}
    \end{center}
    \vspace{-0.5cm}
    \caption{
    (Color) Fitted invariant values for the horizontal (a.) and vertical (b.) planes. 512 random samples are used for each BPM. Samples for each BPM are grouped by vertical lines. The true values of invariants are shown with black dashed line, the blue lines show estimated invariants values with corresponding uncertainties.}
    \label{fig:actionfit}
\end{figure}

For linear motion, two invariants, $I_x$ and $I_y$, exist, also referred to as actions.
To determine the coupled normalization matrix, the objective function below can be minimized at each BPM:
\begin{equation}
    \argmin_{N, I_x, I_y} \sum_i {\left(Q_{x,i}^2 + P_{x,i}^2 - 2 I_x\right)}^2 + {\left(Q_{y,i}^2 + P_{y,i}^2 - 2 I_y\right)}^2
\end{equation}
$(Q_{x,i},P_{x,i},Q_{y,i},P_{y,i})^T = N^{-1} (q_{x,i},p_{x,i},q_{y,i},p_{y,i})^T$ are the normalized coordinates.
Here, $(q_{x,i},p_{x,i},q_{y,i},p_{y,i})$ are TbT coordinates  and reconstructed momenta at the turn $i$, and $N$ is a special normalization matrix introduced in Section~\ref{sec:normal}.
The values of the invariants $I_x$ and $I_y$ are also obtained from the optimization.
Multiple TbT measurements can be utilized as input data.
Similar to the method described in Section~\ref{sec:matrix}, several random samples are used for optimization. 
Figure~\ref{fig:actionfit} shows a model example results for linear invariants.
Focusing errors were introduced to the VEPP-4M lattice, resulting in an average beta function beating of 20\%, and random noise with a standard deviation of \SI{10}{\micro\metre} was added into the TbT signals.


For each monitor, the values and variance of the invariants can be estimated, and a weighted average can be computed to estimate invariants over all BPMs.
This allows to fix the invariant values.
Another option is to use invariant values estimated by using measured amplitudes and either modeled or measured beta functions.
Thus, to fit the elements of the matrix $N$ using fixed invariants $\hat{I}_x$ and $\hat{I}y$, the following objective function needs to be minimized:
\begin{equation}
    \argmin_{N} \sum_i {\left(Q_{x,i}^2 + P_{x,i}^2 - 2 \hat I_x\right)}^2 + {\left(Q_{y,i}^2 + P_{y,i}^2 - 2 \hat I_y\right)}^2
\end{equation}
The minimization procedure is similar to that in the case with free invariants.
As a result, the elements of the normalization matrix and their variances can be estimated.
Using the estimated values and errors, coupled Twiss parameters can be obtained in any required form.


\section{Experimental Results at the VEPP-4M}
\label{sec:vepp}


\begin{figure}[!ht]
    \begin{center}
    \includegraphics[width=\columnwidth,height=0.4\columnwidth]{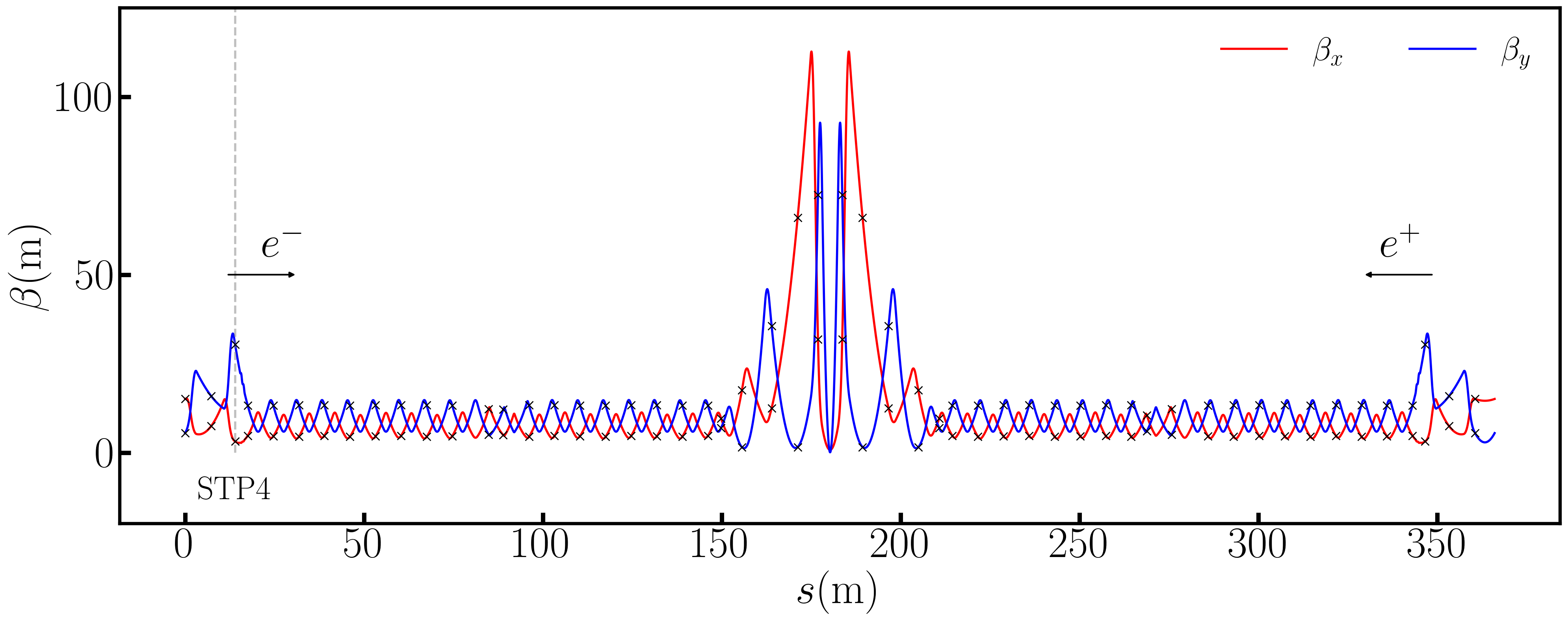}
    \end{center}  
    \vspace{-0.5cm}
    \caption{(Color) The VEPP-4M model beta functions and BPM locations. The beta function values at BPMs are denoted with black crosses.}
    \label{fig:vepp}
\end{figure}

\begin{figure}[!ht]
    \begin{center}
    \includegraphics[width=\columnwidth,height=0.62\columnwidth]{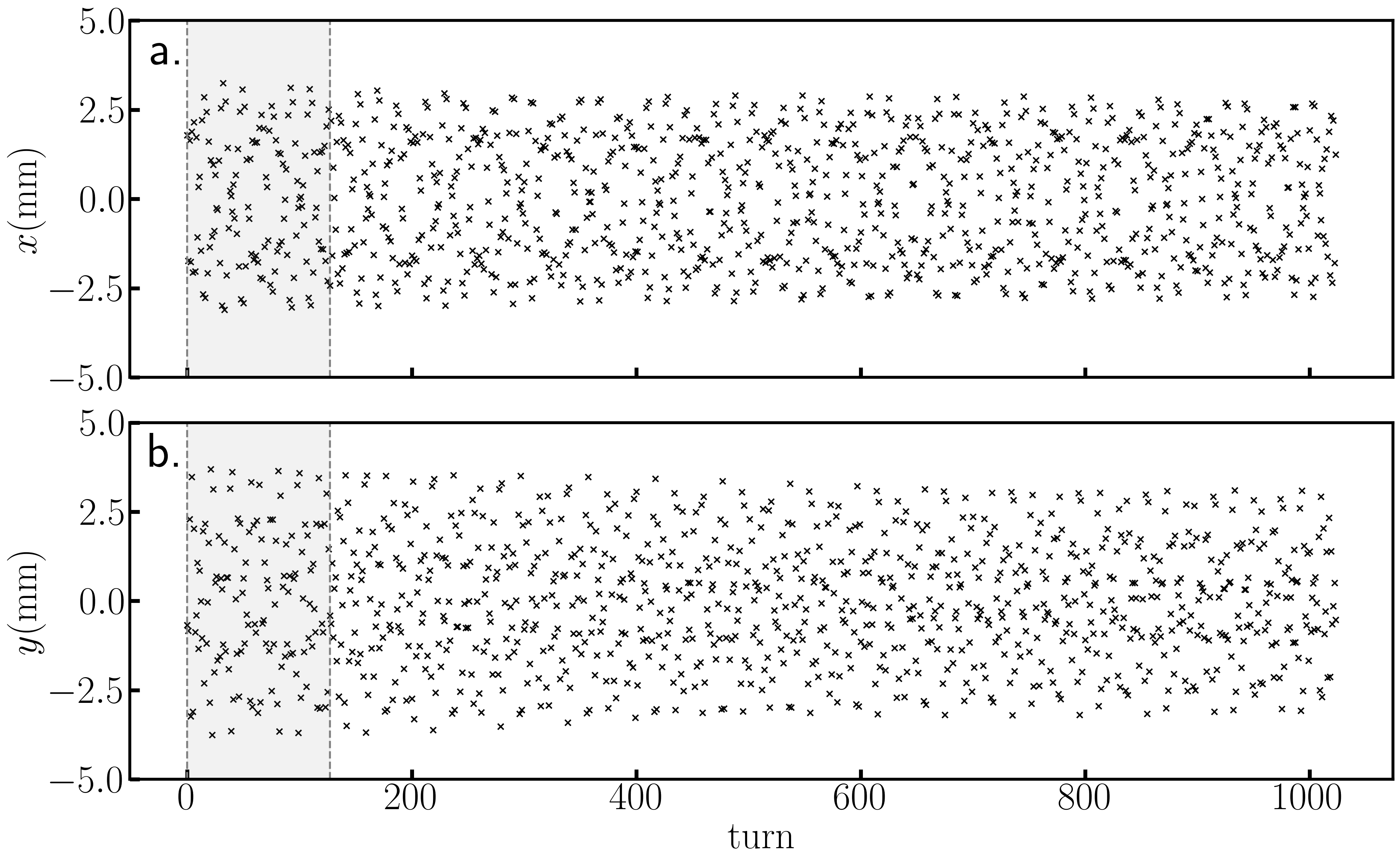}
    \end{center}  
    \vspace{-0.6cm}
    \caption{(Color) Example of measured TbT signals for horizontal (a.) and vertical (b.) planes obtained at STP4 monitor.}
    \label{fig:tbt}
\end{figure}

TbT measurements were performed at the VEPP-4M to test the methods for the estimation of coupled Twiss parameters.
The VEPP-4M is equipped with $54$ dual-plane BPMs (see Figure~\ref{fig:vepp}), capable of recording up to $8192$ turns with a resolution close to \SI{10}{\micro\metre}.
To enhance the effect of coupling, a skew quadrupole in the IP region was used.
Coherent oscillations were excited using a pulsed electromagnetic kicker~\cite{kick}.
The measurements were performed with small positive chromaticity to reduce decoherence effects.
Figure~\ref{fig:tbt} shows an example of TbT signals used in computations.


\vspace{-0.6cm}
\subsection{Uncoupled Twiss Parameters}

\begin{figure}[!ht]
    \begin{center}
    \includegraphics[width=\columnwidth,height=0.4\columnwidth]{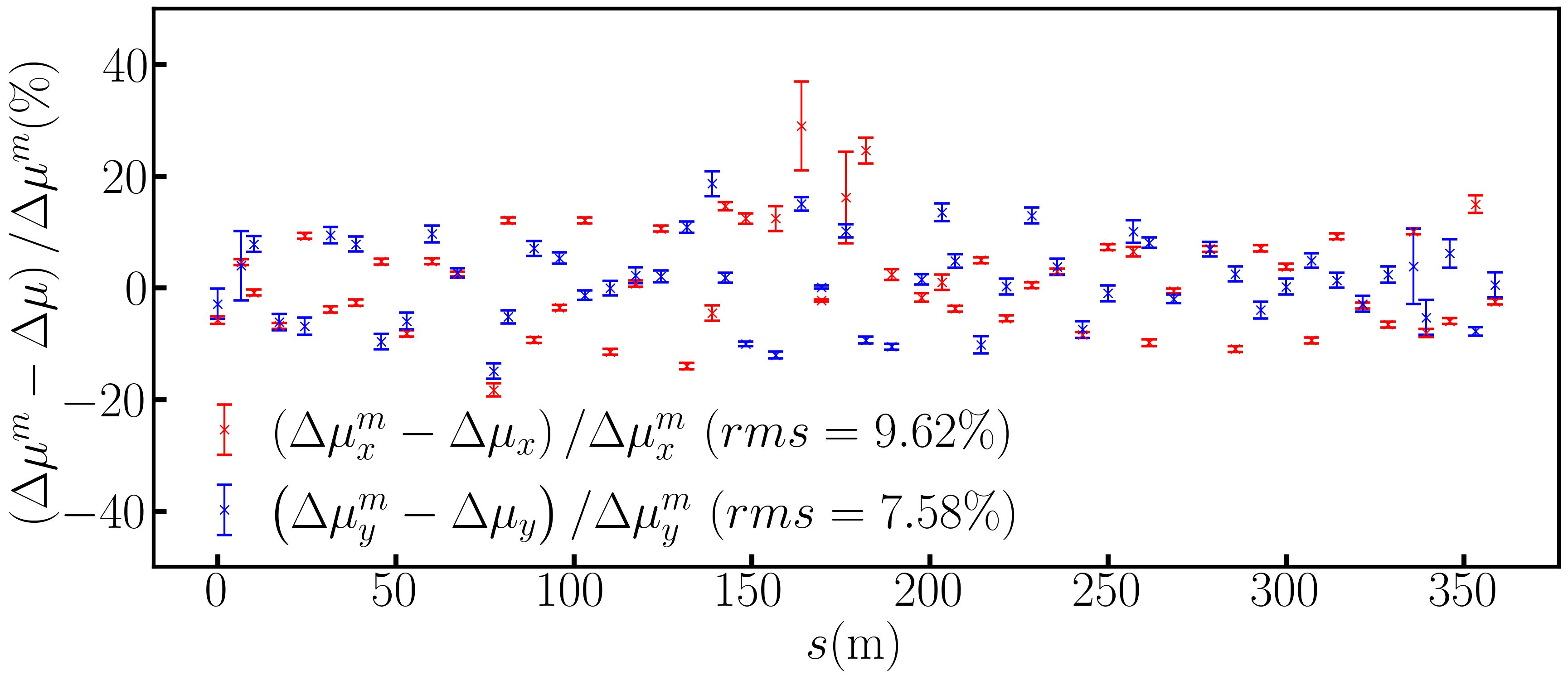}
    \end{center}  
    \vspace{-0.6cm}
    \caption{(Color) Comparison of the measured phase advance between adjacent BPM pairs with corresponding model values $(\Delta \mu^{\textrm{m}} - \Delta \mu) / \Delta \mu^{\textrm{m}}$ for the horizontal (in red) and vertical (in blue) planes.}
    \label{fig:advance}
\end{figure}
\begin{figure}[!ht]
    \begin{center}
    \includegraphics[width=\columnwidth,height=0.62\columnwidth]{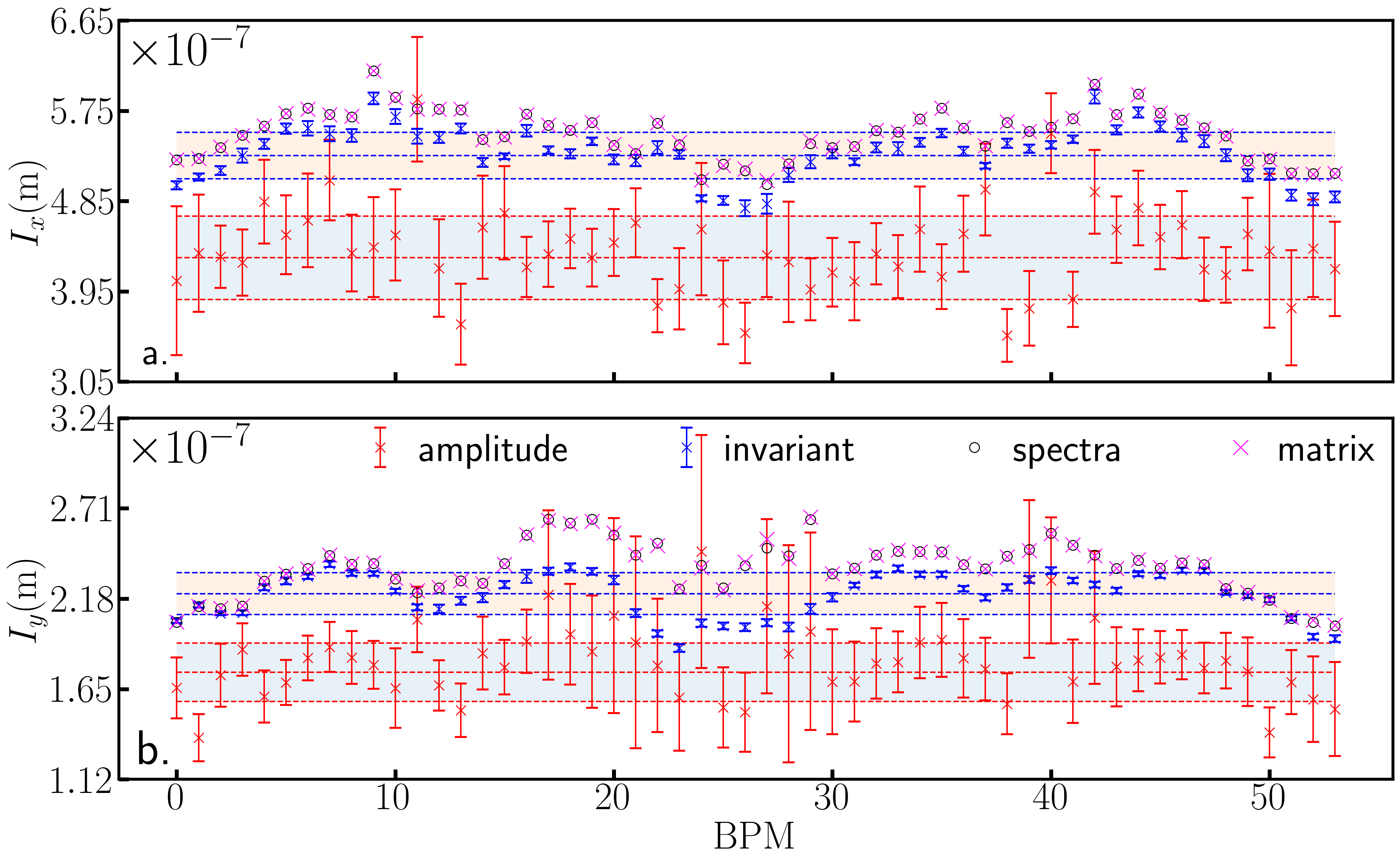}
    \end{center}  
    \vspace{-0.6cm}
    \caption{
    (Color) Estimated horizontal (a.) and vertical (b.) linear invariants for each BPM from TbT amplitudes (in red), invariant fit (in blue), complex normalized spectrum (in black), and reconstructed $n$--turn matrix (in magenta). Corresponding weighted averages with errors are shown with dashed lines for values estimated from amplitude and invariant fit.
    }
    \label{fig:action}
\end{figure}

In Figure \ref{fig:advance}, the measured and model phase advances between adjacent BPMs are shown.
The measured phase advances were computed from TbT data using 512 turns. 
Amplitudes were computed using the first 128 turns, where the effect of decoherence can be neglected (Figure~\ref{fig:tbt}). 
Using the amplitudes and measured beta functions, estimates of linear invariants were obtained at each BPM.
Figure \ref{fig:action} shows the invariant values at each BPM and the corresponding weighted average values (in red).


\begin{figure}[!ht]
    \begin{center}
    \includegraphics[width=\columnwidth,height=0.62\columnwidth]{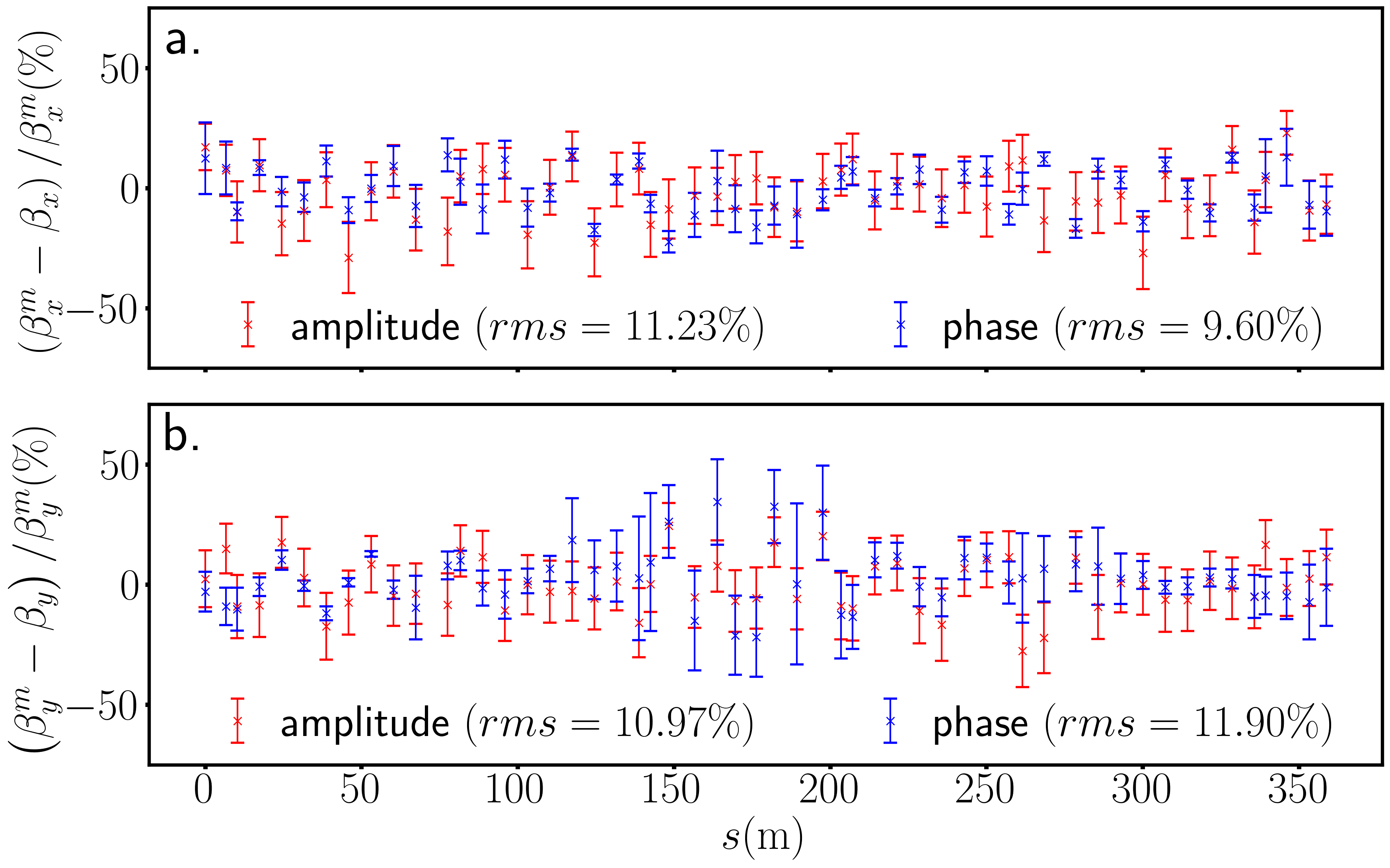}
    \end{center}  
    \vspace{-0.6cm}
    \caption{(Color)  Comparison of the measured beta functions estimated from amplitude (in red) and phase (in blue)  with corresponding model values $(\beta^{\textrm{m}} - \beta) / \beta^{\textrm{m}}$ for horizontal (a.) and vertical (b.) planes. }
    \label{fig:uncoupled}
\end{figure}


Uncoupled Twiss parameters were computed using the amplitudes and phases of TbT signals.
An average over several combinations, with outlier cleaning, is used to compute Twiss parameters from phase data~\cite{rupac2021_twiss}.
The comparison with model values is shown in Figure \ref{fig:uncoupled}.
The discrepancy between measured and model values is around 10\% (rms).
Values obtained by different methods also differ by around 10\% (rms).
Figure \ref{fig:trj} shows an example of phase space trajectory normalization using model Twiss parameters (in black) and measured ones estimated phase data (in red).
In both cases horizontal and vertical trajectories deviate from circles.
To perform full normalization, coupling should be included.


\subsection{Coupled Twiss Parameters}

\begin{figure}[!ht]
    \begin{center}
    \includegraphics[width=\columnwidth,height=1.23\columnwidth]{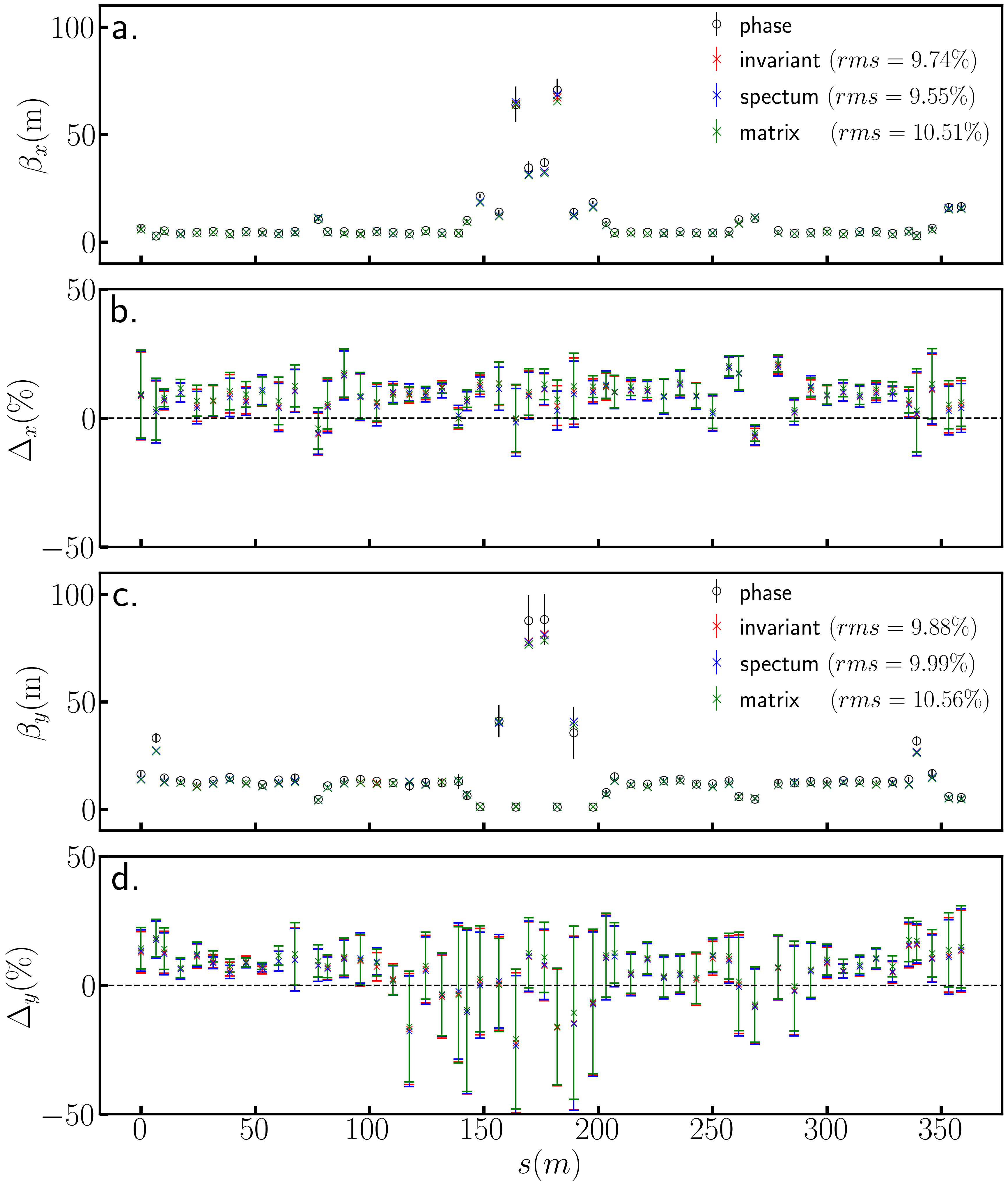}
    \end{center}  
    \vspace{-0.5cm}
    \caption{(Color) Comparison of in-plane beta function estimated using different methods based on momentum reconstruction for horizontal (a.) and vertical (c.) planes. Corresponding deviations from the beta functions obtained using phase data $\Delta = (\beta^{\varphi} - \beta) / \beta^{\varphi}$ are shown in (b.) and (d.).}
    \label{fig:twiss}
\end{figure}

Figure \ref{fig:twiss} shows the estimated in-plane beta functions obtained using methods described in Sections \ref{sec:matrix}, \ref{sec:spectrum}, and \ref{sec:invariant}.
For comparison, beta functions estimated using the method based on TbT phases are also shown.
Figure \ref{fig:twiss} (b. and d.) specifically shows the deviation of in-plane beta functions relative to the values estimated based on phases.
All methods for estimating coupled Twiss parameters use reconstructed momentum, which was computed from LS fit using two monitors around the probed monitor and modeled transport matrices.
In all cases, the first 128 turns were used from a single TbT measurement.
All methods demonstrate similar results when compared to the method based on phases.
This agreement also indicates that the Twiss parameters specific to coupling, estimated using these methods, are expected to be meaningful and can be used for coupling characterization.
Estimated linear invariants based on different methods are shown in Figure~\ref{fig:action}.


Estimated uncertainties for the results obtained with the $M^k$ fit method are several times smaller than those for the other two methods.
This is related to the optimal selection of $k$ power, as described in~\cite{robust}. 
The optimal value of $k$ for these experimental data is 10.
If computed with a non-optimal value, for example, $k=1$, the values of uncertainties rise to those of the other methods.
Additionally, at the IP BPMs, the estimated Twiss parameters are significantly off, exhibiting large uncertainties.
The symplectified matrix norm errors, $|\hat M^k - M^k|/|M^k|$, are around 5\% (rms).


\begin{figure}[!ht]
    \begin{center}
    \includegraphics[width=\columnwidth,height=0.49\columnwidth]{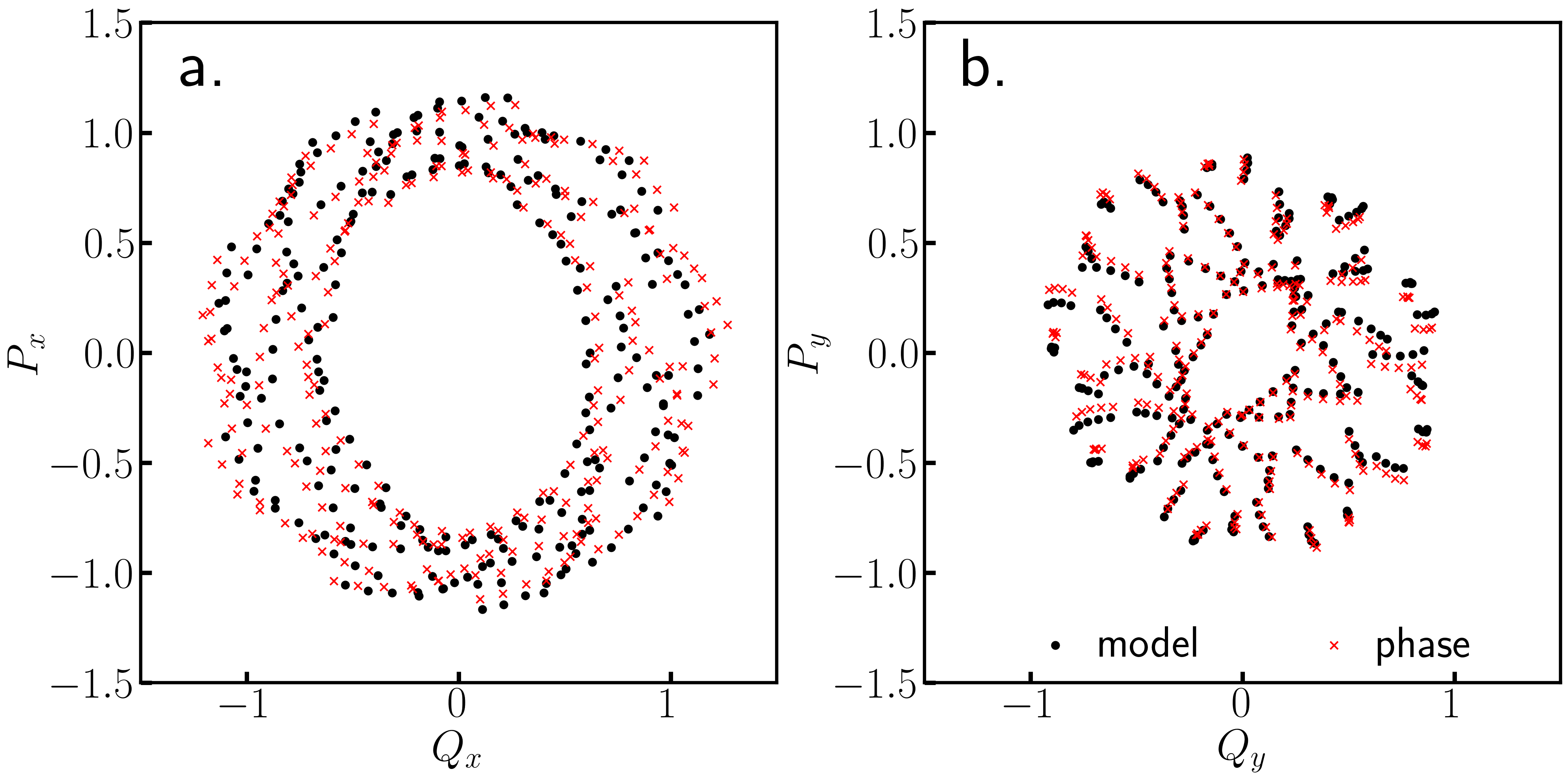}
    \includegraphics[width=\columnwidth,height=0.49\columnwidth]{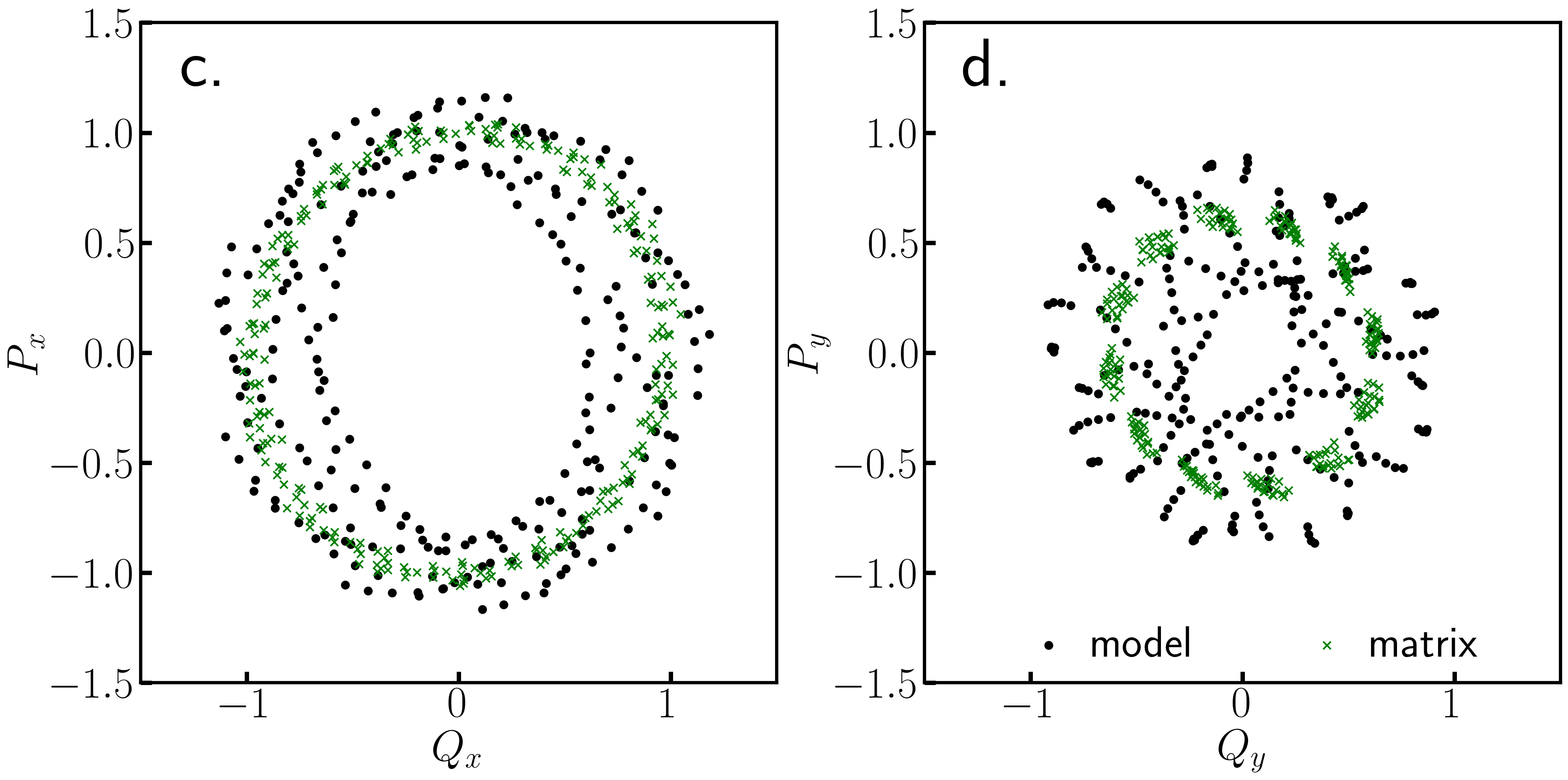}
    \includegraphics[width=\columnwidth,height=0.49\columnwidth]{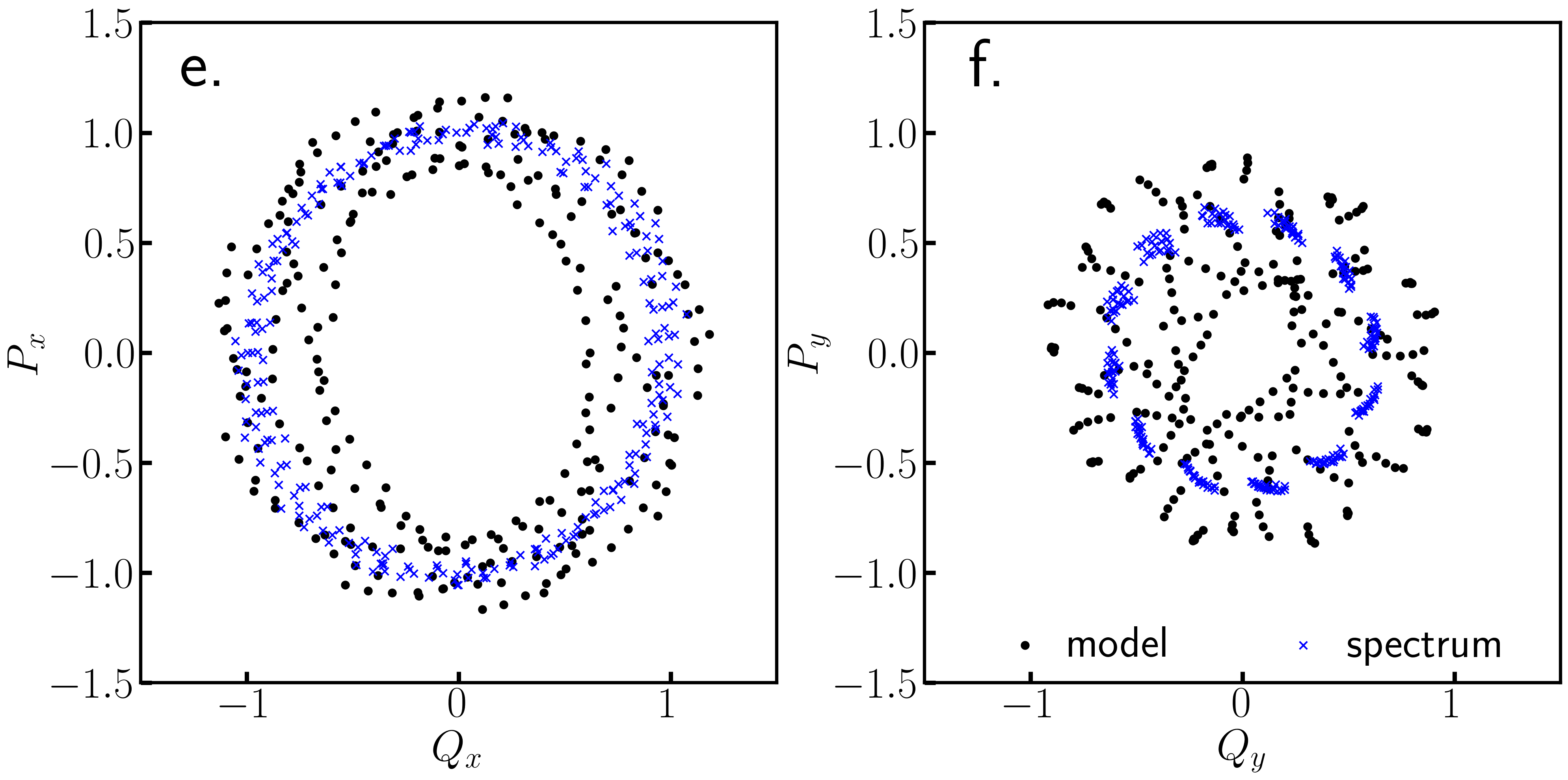}
    \includegraphics[width=\columnwidth,height=0.49\columnwidth]{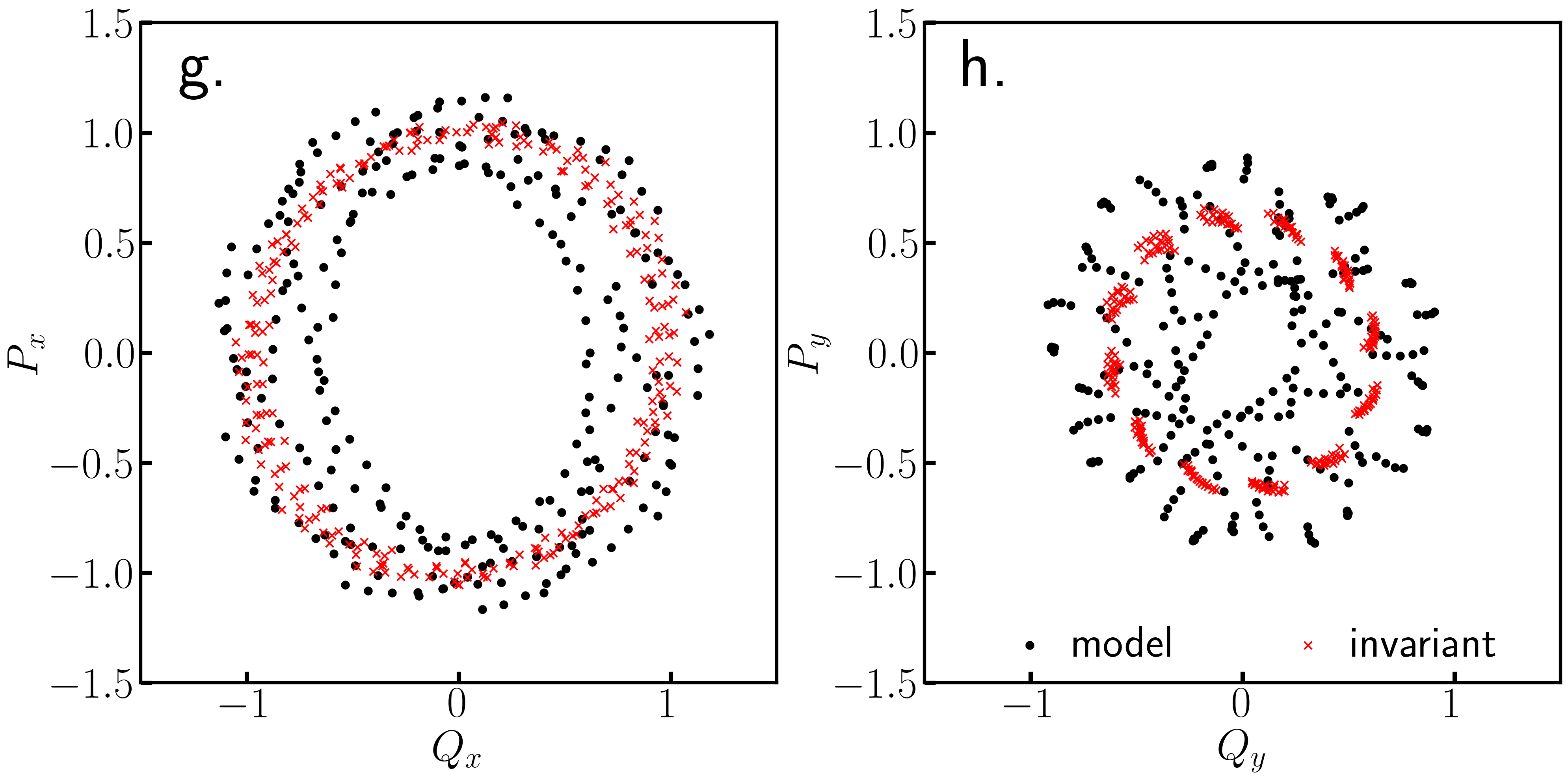}
    \end{center}  
    \vspace{-0.5cm}
    \caption{(Color) Normalized phase space trajectory (first 256 turns are shown) at STP4 monitor computed using the model normalization matrix compared with the trajectory using the measured normalization matrix based on phase data (a. and b.), reconstruction of the one--turn matrix (c. and d.),  complex normalized spectrum (e. and f.), and coupled linear invariants (g. and h.).}
    \label{fig:trj}
\end{figure}


In general, all methods are in good agreement when comparing estimated values of Twiss parameters.
For example, the covariance matrix constructed from the values of in-plane beta functions obtained by different methods has all its elements close in magnitude, which indicates a similar variation of the results obtained with different methods. Figure \ref{fig:action} shows invariant values computed using the estimated normalization matrix, which is based on the complex normalized spectrum.
In normalized coordinates, $I = 1/2 \left( Q^2 + P^2 \right)$ is a linear invariant that can be transformed to the laboratory frame using Equation~\ref{frame}.
For the invariant fit method, the values of invariants are directly fitted.
The coupled invariants, computed by different methods, are in agreement.
Also, it can be noted that these invariants differ from the uncoupled estimation.


Using the full set of coupled Twiss parameters, one can build a coupled normalization matrix and use it to normalize measured phase space trajectories.
Figure \ref{fig:trj} shows an example of such normalization using both uncoupled and coupled normalization matrices. 
In the case of the uncoupled normalization matrix, there is a strong deviation from a circle due to introduced coupling.
In all cases, the shape of the horizontal phase space trajectory is close to a circle after coupled normalization.
The latter is further evidence that the considered methods produce sensible results.
For the $M^k$ fit method, the result for the vertical plane (Figure~\ref{fig:trj} d.) has more residual smear compared to other two methods.
Overall, the normalization performance for all methods is much better than that of the model or the method based on TbT phases.

\begin{figure}[!ht]
    \begin{center}
    \includegraphics[width=\columnwidth,height=0.62\columnwidth]{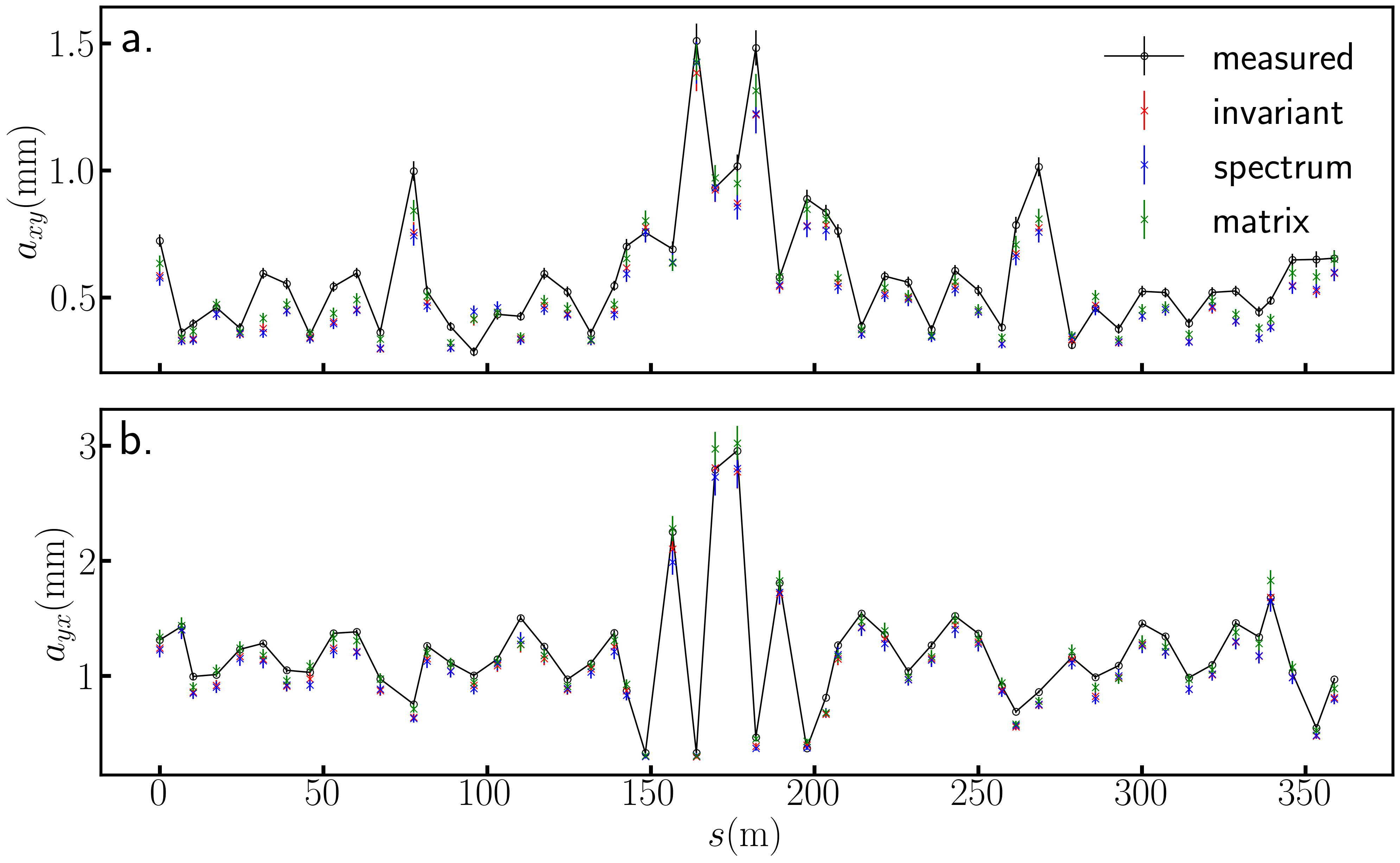}
    \end{center}  
    \vspace{-0.5cm}
    \caption{
    (Color) Comparison of coupled amplitudes for horizontal (a.) and vertical (b.) planes, computed directly from TbT data (in black) and using estimated linear invariants (from fit) together with coupled Twiss parameters estimated from fit (in red), complex normalized spectrum (in blue), and reconstructed one--turn matrix (in green).
    }
    \label{fig:amplitude}
\end{figure}

Another option to check the validity of the estimated coupled Twiss parameters is to compute the coupled amplitudes $a_{x, y}$ and $a_{y, x}$ defined in Section~\ref{sec:spectrum}.
These amplitudes can be computed directly from TbT data and using estimated linear invariants and coupled Twiss parameters as follows:
\begin{align}
    & a_{x, y}^2 = 2 \hat I_y \left(n_{1,3}^2 + n_{1,4}^2\right) \nonumber\\
    & a_{y, x}^2 = 2 \hat I_x \left(n_{3,1}^2 + n_{3,2}^2\right)
\end{align}
where $n_{i,j}$ are the elements of the normalization matrix defined in Section~\ref{sec:normal}.
The result of this comparison is shown in Figure~\ref{fig:amplitude}.
A qualitative argument for the validity of all methods can be observed, which further indicates the correctness of the estimated coupled Twiss parameters.


\section{Summary}
\label{sec:summary}


In this paper, we examine the theoretical foundations of several methods for estimating the Twiss parameters, taking into account the coupling of transverse oscillations.
These methods provide complete normalization matrices at each BPM.
Methods for the estimation of coupled Twiss parameters can be accessed in the reference~\cite{harmonica}.
In all methods, the reconstructed momenta, which can be computed using several BPMs and either modeled or measured transport matrices between them, are used.
Since the methods directly use TbT coordinates, they are sensitive to BPM calibration errors.


All methods allow for the construction of full normalization matrices at all BPMs and do not rely on the weak coupling assumption.
These matrices can be used to perform linear normalization of TbT data, compute local transport matrices, and propagate coupled Twiss parameters using model transport matrices.
Estimations of linear coupled invariants are available either directly from the fit itself or via estimated normalization matrices.
Statistical methods are used to estimate uncertainties due to random variations in TbT data.


The method based on reconstructing the one-turn matrix or its power uses TbT coordinates and reconstructed momenta to fit the corresponding matrix.
Symplectification procedure is performed on the estimated matrix prior to the computation of coupled Twiss parameters.
Another method uses normalized complex spectra and a special form of the normalization matrix to minimize an objective that depends on coupling amplitudes.
This minimization procedure provides an estimation of the coupled normalization matrix.


In the last method, the objective function to be minimized involves linear coupled invariants.
These invariants can be estimated from the minimization procedure, fixed by prior optimization, or computed by other means.
Numerical simulations show high accuracy in determining the coupled Twiss parameters using the described methods, taking into account random BPM errors and model errors.


Experimental verification of the methods was carried out at the VEPP-4M accelerator, showing good agreement between measurements obtained by different methods.
The measured in-plane beta functions are close to those obtained using TbT phase data.
Using the measured normalization matrices, the normalized phase space trajectories are close to circles for both transverse planes, and the estimated coupling amplitudes qualitatively match the directly measured ones. 


Based on experimentally estimated Twiss parameters, it can be concluded that the methods have similar performance. 
The method based on the one-turn matrix fit, or its power, has smaller uncertainties when used with optimal value of power, but also exhibits more residual smear, as evident from the normalization of the phase trajectories. 
This method, along with the method based on the fit of coupled invariants, can combine data from several TbT measurements, for example, when the usable data length is limited.
The method based on the complex normalized spectrum uses a single set of TbT data measurements.
The structure of the objective function used in this method makes it less sensitive to TbT signals decoherence, since it utilizes ratios of harmonic amplitudes.


Several tests were performed to assess the validity of the estimated coupled Twiss parameters: comparison with beta functions computed based on phase data, estimation of coupled actions, normalization of phase space trajectories, and reconstruction of coupled amplitudes.
The results of these tests allow us to conclude that the methods described provide an adequate description of the coupled transverse motion and can be used for its characterization.


\section{Acknowledgments}

This work was partially supported by the Ministry of Science and
Higher Education of the Russian Federation within the governmental order for
Boreskov institute of Catalysis (project FWUR-2024-0041).





%

\end{document}